\PassOptionsToPackage{hyphens}{url}

\documentclass[lettersize,journal]{IEEEtran}
\usepackage{microtype}
\usepackage{subfig, siunitx, adjustbox}
\usepackage{booktabs} 

\usepackage{stfloats}
\usepackage{url}
\usepackage{array}
\usepackage{mathtools}
\usepackage{amsthm}
\usepackage{adjustbox}
\usepackage{algpseudocode}
\usepackage{algorithm}
\usepackage{wrapfig}
\usepackage{multirow, multicol}
\usepackage[table]{xcolor}

\usepackage{color,soul}

\definecolor{hcolor}{rgb}{0.4,0.2,0.8} 
\sethlcolor{green} 

\DeclareMathOperator*{\argmax}{arg\,max}

\usepackage{cite}
\usepackage{amsmath,amssymb,amsfonts}
\usepackage{graphicx}
\usepackage{textcomp}
\usepackage{xcolor}
\def\BibTeX{{\rm B\kern-.05em{\sc i\kern-.025em b}\kern-.08em
    T\kern-.1667em\lower.7ex\hbox{E}\kern-.125emX}}

\makeatletter
\AfterEndEnvironment{algorithm}{\let\@algcomment\relax}
\AtEndEnvironment{algorithm}{\kern2pt\hrule\relax\vskip3pt\@algcomment}
\let\@algcomment\relax
\newcommand\algcomment[1]{\def\@algcomment{\footnotesize#1}}

\renewcommand\fs@ruled{\def\@fs@cfont{\bfseries}\let\@fs@capt\floatc@ruled
  \def\@fs@pre{\hrule height.8pt depth0pt \kern2pt}%
  \def\@fs@post{}%
  \def\@fs@mid{\kern2pt\hrule\kern2pt}%
  \let\@fs@iftopcapt\iftrue}
\makeatother

\theoremstyle{plain}

\theoremstyle{definition}

\theoremstyle{remark}

\usepackage{hyperref}

\usepackage[compress, numbers]{natbib}

\begin{document}

\title{Standards-Compliant DM-RS Allocation via Temporal \\ Channel Prediction for Massive MIMO Systems
}


\author{Sehyun Ryu and Hyun Jong Yang,~\IEEEmembership{Senior Member,~IEEE}
\thanks{Sehyun Ryu is with the Department of Electrical Engineering, Pohang University of Science and Technology (POSTECH), Pohang, Republic of Korea (e-mail: sh.ryu@postech.ac.kr).}
\thanks{Hyun Jong Yang is the corresponding author and is with the Department of Electrical and Computer Engineering and the Institute of New Media and Communications, Seoul National University (SNU), Seoul, Republic of Korea (e-mail: hjyang@snu.ac.kr).}}



\maketitle

\begin{abstract}
Reducing CSI feedback overhead in beyond 5G networks is a critical challenge.
The growing number of antennas in modern massive MIMO systems substantially increases the channel state information (CSI) feedback demand in frequency division duplex (FDD) systems.
To address this, extensive research has focused on CSI compression and prediction, with neural network-based approaches gaining momentum and being considered for integration into the 3GPP 5G-Advanced standards.
While deep learning has been effectively applied to CSI-limited beamforming and handover optimization, reference signal allocation under such constraints remains comparatively underexplored.
To fill this gap, we introduce the concept of channel prediction-based reference signal allocation (CPRS), which jointly optimizes channel prediction and DM-RS allocation to improve data throughput without requiring CSI feedback.
We further propose a standards-compliant ViViT/CNN-based architecture that implements CPRS by treating evolving CSI matrices as sequential image-like data.
This design enables efficient and adaptive transmission in dynamic environments.
Ray-tracing-based simulations in NVIDIA Sionna validate the proposed method, demonstrating up to 36.60\% throughput improvement over benchmark strategies.  
\end{abstract}

\begin{IEEEkeywords}
5G NR, Massive MIMO, Reference Signal, DM-RS, Channel Prediction, Deep Learning.
\end{IEEEkeywords}

\section{Introduction}
Multiple-input multiple-output (MIMO) systems \cite{GESBERT2002} have been at the core of wireless innovation since 4G LTE.
However, the introduction of massive MIMO has led to the challenge of increased channel dimensionality \cite{LU2014}.
As the dimension of the channel matrix increases, both the transmission of the reference signal for channel estimation and the feedback of downlink channel state information (CSI) to the base station have become significant sources of overhead.
The downlink CSI is critical for enhancing communication operations such as \textit{beamforming} \cite{JAFAR2001}, \textit{handover} \cite{TOLLI2008}, and \textit{reference signal allocation} \cite{CAI2005}.
In time division duplexing (TDD) systems, the downlink channel can be inferred from the uplink channel due to channel reciprocity. 
However, in frequency division duplexing (FDD) systems, CSI feedback is required \cite{LOVE2008}.
As CSI feedback has become a major source of overhead in FDD systems, extensive research has been conducted to address this challenge \cite{SHEN2017}. 

\begin{figure}[t]
\vspace{0pt}
    \centering
    \includegraphics[draft=false, width= 0.82 \linewidth]{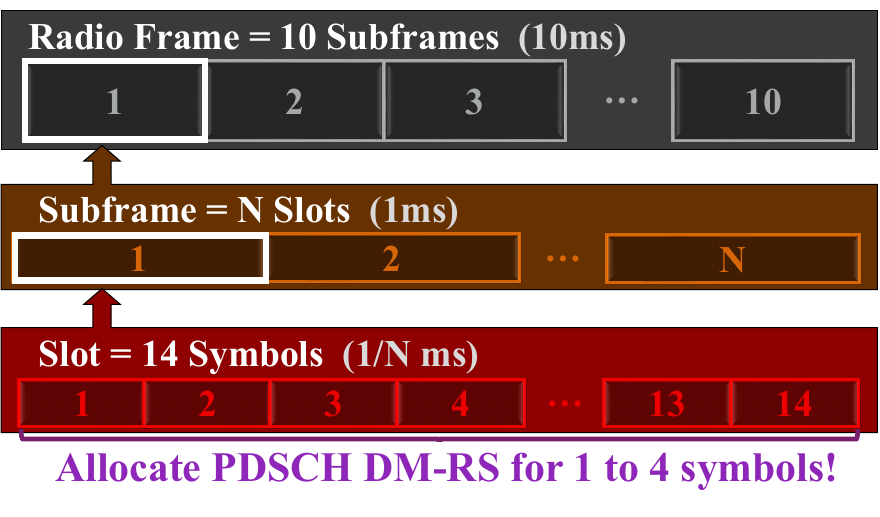}
\vspace{-7pt}
    \caption{The NR frame structure and its corresponding transmission timings are defined based on a 10ms radio frame, consisting of ten 1ms subframes. The number of slots per subframe, N, is flexibly determined by the numerology specified in 3GPP TS 38.211 table 4.3.2-1 \cite{TS138211}, where $\textnormal{N} = 2^{\mu}$, $\mu \in \mathbb{Z}$, and $0 \leq \mu \leq 6$. Within each slot, the PDSCH DM-RS can be configured to occupy 1 to 4 out of the 14 OFDM symbols.}
    \label{fig:5G_FrameStructure}
    \vspace{-14pt}
\end{figure}

Recently, deep learning has emerged as a promising solution for reducing CSI feedback overhead and has been actively discussed for incorporation into 5G-Advanced standards following 3GPP Rel-18 \cite{RP213599}.
The solutions can be broadly classified into two categories: CSI compression \cite{WEN2018} and channel prediction \cite{WANGJIE2019}.
We focus on channel prediction, which estimates downlink CSI from uplink CSI or data signals, even without explicit feedback.
The network architectures have evolved from convolutional neural networks (CNNs) \cite{WANGJIE2019} to long short-term memory (LSTM) models \cite{HELMY2023}, and more recently, transformers \cite{ZHOU2024}.
There have also been attempts to jointly optimize beamforming and handover.
For instance, \cite{CHU2022} proposed a deep reinforcement learning algorithm for channel prediction and beamforming, while \cite{JIANG2022} used a transformer-based approach for beamforming with predicted CSI.
Channel quality prediction has also been used to guide handover decisions \cite{SKABA2024}.
However, to the best of the authors' knowledge, no prior study has addressed the critical issue of reference signal allocation in conjunction with channel prediction, despite their strong interdependence.

We address the allocation of the physical downlink shared channel (PDSCH) demodulation reference signal (DM-RS) within the NR frame structure, as illustrated in Fig.~\ref{fig:5G_FrameStructure}.
\IEEEpubidadjcol Among various reference signals defined in 3GPP NR, the PDSCH DM-RS plays a critical role in determining data throughput, as it directly enables the user equipment (UE) to estimate the channel for data demodulation.
Moreover, unlike other reference signals transmitted over the control plane, the DM-RS is carried on the data plane alongside data symbols.
The NR specification supports 1–4 DM-RS symbols per slot with multiple time-domain patterns (see Table~\ref{tab:TS38211}), offering flexibility to accommodate mobility, noise, and channel estimation uncertainty.
The in-slot placement makes DM-RS allocation particularly important.
Increasing the number of DM-RS symbols improves channel estimation accuracy but reduces the number of transmittable data symbols, highlighting an inherent trade-off in the allocation design.
Therefore, the DM-RS must be optimally allocated to balance channel estimation quality and the number of available data symbols, considering the current downlink CSI.
However, as the standard does not specify which allocation to use under specific downlink CSI conditions, network operators have the flexibility to configure it based on their own channel assessments.

\begin{table}[t]
\centering
\caption{DM-RS positions within a 14-symbol slot duration are specified in Table 7.4.1.1.2-3/4 of 3GPP TS 38.211 \cite{TS138211}, with $l_0 \in \{2, 3\}$ and $l_1 \in \{11, 12\}$. Only Type-A configurations are included, as Type-B applies to a special use case of mini-slot-based scheduling with fewer supported allocation patterns.} 


\label{tab:TS38211}
    \adjustbox{width=0.92\linewidth}
    {
        \begin{tabular}{w{c}{1.5cm} w{c}{1cm} w{c}{1cm} w{c}{1cm} w{c}{1cm}} 
            \toprule[1pt]
            \multirow{2}{*}[-4pt]{DM-RS Length} & \multicolumn{4}{c}{DM-RS positions} \\
             \cmidrule[0.5pt](l{1pt}r{1pt}){2-5}
              & pos0 & pos1 & pos2 & pos3\\
            \cmidrule[0.5pt](l{1pt}r{1pt}){1-5}
            Single-Symbol & $l_{0}$   & $l_{0}$, $l_{1}$  & $l_{0}$, 7, 11    & $l_{0}$, 5, 8, 11 \\
            \cmidrule[0.5pt](l{1pt}r{1pt}){1-5}
            Double-Symbol & $l_{0}$   & $l_{0}$, 10  & -   & - \\
            \bottomrule[1pt]
        \end{tabular}

    }
    \vspace{-8pt}
\end{table}

This paper proposes a joint optimization framework for channel prediction and reference signal allocation, with a particular focus on DM-RS, instead of treating them as independent modules.
Although motivated by B5G scenarios, the proposed approach is fully compatible with the 5G NR frame structure and applicable under current standard specifications.

\textbf{Contribution:}
1) This paper introduces \textcolor{hcolor}{the concept of channel prediction-based reference signal allocation (CPRS)}, focusing on DM-RS as a practical means of reducing CSI feedback overhead in massive MIMO FDD scenarios.
2) We propose \textcolor{hcolor}{a video vision transformer (ViViT)/CNN-based~\cite{ARNAB2021} CPRS algorithm, designed with consideration of the NR frame structure}, which interprets time-varying uplink channel matrices as video data.
3) We \textcolor{hcolor}{generate ray-tracing-based channel data and perform simulations using NVIDIA Sionna~\cite{NVIDIA}}, demonstrating the effectiveness of the proposed method.

\section{System Model and Problem Formulation}

We consider a massive MIMO orthogonal frequency division multiplexing (OFDM) communication system, focusing on a downlink scenario with UE mobility, where the BS and UE communicate based on the NR frame structure.
The numbers of receive antennas, transmit antennas, and subcarriers are denoted by $N_{\textnormal{R}}$, $N_{\textnormal{T}}$, and $N_{\textnormal{C}}$, respectively.
The antennas at the UE receive the same transmitted signal, and the received signal for the $t$-th OFDM symbol in the slot and $k$-th subcarrier can be expressed as follows:
\begin{align}
    y_{t,k} = \mathbf{h}_{\textnormal{DL},t,k} \, \mathbf{v}_{t,k} \, x_{t,k} + \mathbf{w}_{t,k},
\end{align}
where $\mathbf{h}_{\textnormal{DL},t,k} \in \mathbb{C}^{N_{\textnormal{R}} \times N_{\textnormal{T}}}$ is the downlink channel matrix associated with the $t$-th symbol and $k$-th subcarrier, $\mathbf{v}_{t,k} \in \mathbb{C}^{N_{\textnormal{T}} \times 1}$ is the precoding vector, $x_{t,k} \in \mathbb{C}$ is the transmit symbol, and $\mathbf{w}_{t,k} \in \mathbb{C}^{N_{\textnormal{R}} \times 1}$ denotes additive noise.
Each $t$-th OFDM symbol has a corresponding downlink CSI matrix,
\begin{align}
    \mathbf{H}_{\textnormal{DL},t} = \left[ \mathbf{h}_{\textnormal{DL},t,1} \, \mathbf{h}_{\textnormal{DL},t,2} \, \cdots \, \mathbf{h}_{\textnormal{DL},t,N_{{\textnormal{C}}}} \right] \in \mathbb{C}^{N_{\textnormal{R}} \times N_{\textnormal{T}} \times N_{\textnormal{C}}}.
\end{align}
We define the collection of downlink CSI across the 14 OFDM symbols in a slot as
\begin{align}
    \mathbf{H}_{\textnormal{DL}} = \left\{ \mathbf{H}_{\textnormal{DL},t} \right\}_{t=1}^{14}.
\end{align}


In our system, the BS is assumed to predict $\mathbf{H}_{\textnormal{DL}}$ using the uplink CSI from the current slot, denoted as $\mathbf{H}_{\textnormal{UL}} = \{ \mathbf{H}_{\textnormal{UL},i} \in \mathbb{C}^{N_{\textnormal{T}} \times N_{\textnormal{R}} \times N_{\textnormal{C}}}\}_{i \in I_{u}}$, where $I_{u}$ is the set of DM-RS symbol indices in the current uplink slot.
We assume that the uplink DM-RS is transmitted at four symbols per slot and $I_{u} = \{ 2,5,8,11 \}$.
The downlink channel prediction using a predictor function $f(\cdot)$ is formulated as
\begin{align}
    \hat{\mathbf{H}}_{\textnormal{DL}} = f (\mathbf{H}_{\textnormal{UL}})
    \label{eq:channel_prediction}.
\end{align}

\noindent The BS utilizes $\hat{\mathbf{H}}_{\textnormal{DL}}$ to identify the optimal allocation $p^{*} \in \mathcal{P}$, where $\mathcal{P}$ denotes the set of possible allocation configurations from which the BS selects one to configure the next slot, in order to maximize data throughput.
We refer to this process as \textit{channel prediction-based reference signal allocation (CPRS)}, and define the corresponding optimization problem as follows:
\begin{align}
   p^{*} = \argmax_{p \in \mathcal{P}} \: \left\{ R \cdot \textnormal{D}_{\textnormal{sent}} \cdot \left( 1 - \frac{n_{p}}{n_{s}} \right) \cdot \left( 1 - \textnormal{BLER} \right) \right\},
   \label{eq:optimization_problem}
\end{align}

\begin{figure*}[ht]
\vspace{-10pt}
    \centering
    \includegraphics[draft=false, width= 0.77 \linewidth]{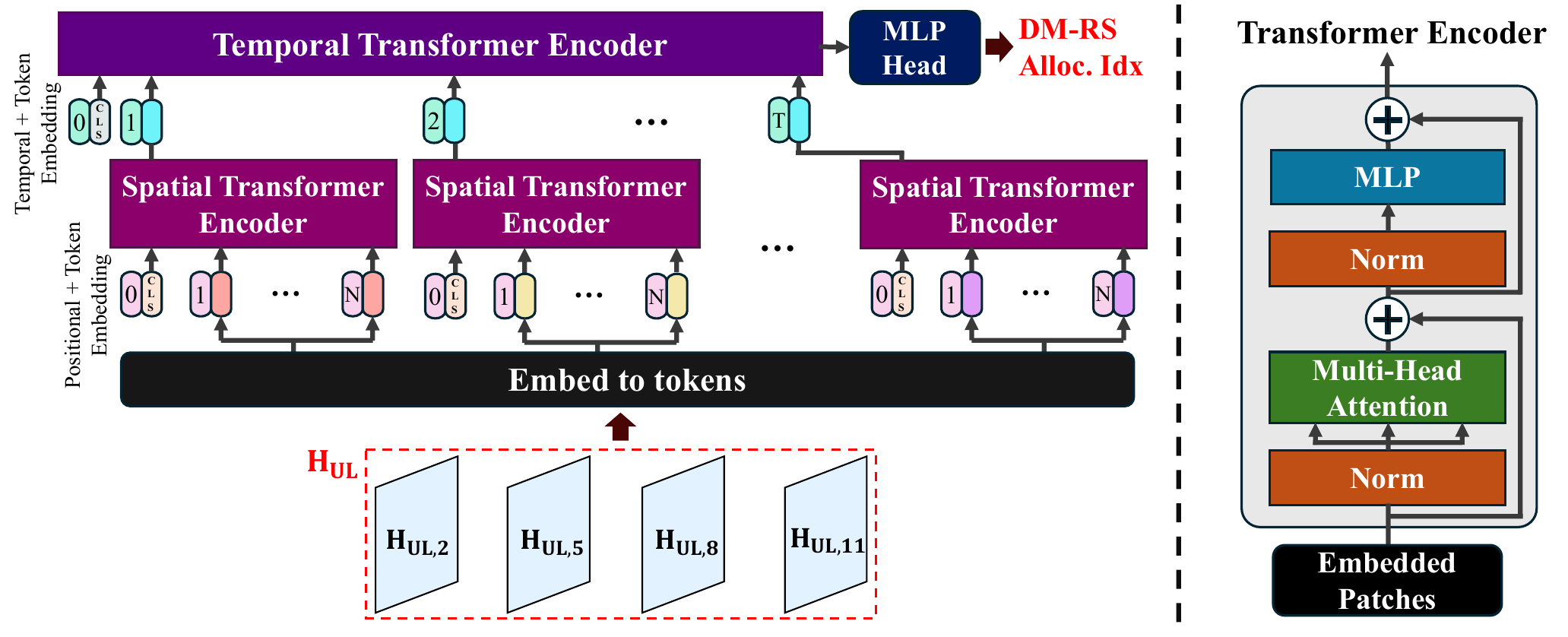}
    \caption{The proposed ViViT-based CPRS network captures the temporal variation of uplink channel matrices $\mathbf{H}_{\textnormal{UL}}$, obtained from DM-RS symbols at positions $I_{u} = \{ 2,5,8,11 \}$ within an uplink slot. Leveraging a transformer architecture optimized for video data, it predicts the optimal DM-RS allocation for the next slot.}
    \label{fig:CPRS_network}
    \vspace{-11pt}
\end{figure*}

\noindent given $\hat{\mathbf{H}}_{\textnormal{DL}} = f(\mathbf{H}_{\textnormal{UL}})$, where $R$ is the code rate, $\textnormal{D}_{\textnormal{sent}}$ is the total number of bits transmitted in one slot, $n_{p}$ is the number of DM-RS symbols in allocation $p$, $n_{s}=14$ is the number of symbols within a slot, and BLER denotes the block error rate at the UE.
The objective function of (\ref{eq:optimization_problem}) represents the data throughput (bits per slot), following the model in \cite{DAS2015}.
For a fixed numerology, $R$, $\textnormal{D}_{\textnormal{sent}}$, and $n_s$ are constants, while $n_p$ is determined by the allocation $p$.
Thus, the main difficulty is to model the relationship among $p$, $\mathbf{H}_{\textnormal{UL}}$, and the resulting BLER at the UE in the next slot.


Even if the BS accurately estimates the $\mathbf{H}_{\textnormal{DL}}$, it remains challenging to determine the extent of BLER caused by channel errors in symbols between DM-RS placements.
We adopt a data-driven approach to determine the relationship among $p$, $\mathbf{H}_{\textnormal{UL}}$, BLER, and ultimately, data throughput.
Leveraging neural network algorithms commonly used for nonconvex optimization, we propose an end-to-end method that jointly optimizes the channel prediction problem in ($\ref{eq:channel_prediction}$) and the DM-RS allocation in ($\ref{eq:optimization_problem}$).
In particular, the proposed CPRS framework directly classifies the optimal allocation $p^{*} \in \mathcal{P}$ from $\mathbf{H}_{\textnormal{UL}}$.
For the neural network-based classifier $\mathcal{F}(\cdot;\Theta)$ and estimated allocation $\hat{p}^* \in \mathcal{P}$, the proposed CPRS method is defined as
\begin{align}
    \hat{p}^{*} = \mathcal{F}(\mathbf{H}_{\textnormal{UL}}; \Theta)
    \label{eq:CPRS_classifier}.
\end{align}
\noindent We assume the possible allocation set $\mathcal{P}$ follows the NR standard in Table \ref{tab:TS38211}.
Without loss of generality, to simplify our analysis in this study, we consider the typical case of $l_{1} = 11$, yielding $|\mathcal{P}| = 13$ possible allocations, and apply Kronecker placement to focus on time-domain symbol allocation, as frequency-domain patterns (Type 1 and Type 2) are limited to only two fixed configurations and offer little flexibility. 

\section{ViViT/CNN-based CPRS Network}
\label{SEC:CPRS_ViViT}

The CPRS framework predicts the optimal DM-RS allocation by exploiting the temporal evolution of the uplink channel matrices $\mathbf{H}_{\textnormal{UL}}$ and is inherently \textit{model-agnostic}, allowing the use of various backbone architectures.
In this paper, we instantiate CPRS using two empirically strong backbones: ViViT and a CNN model.
ViViT provides global spatio–temporal modeling and preserves channel structures that are critical for DM-RS allocation.
The CNN-based variant maintains the same input–output structure while offering lower computational complexity; its standard architecture is omitted for brevity.

Before being processed by the backbone model, each uplink channel matrix 
$\mathbf{H}_{\textnormal{UL},i}$ for $i \in \{2,5,8,11\}$ is decomposed into real and imaginary parts
and reshaped into a unified real-valued input tensor:
\begin{align}
    \mathbf{H}_{\textnormal{UL},i} \in \mathbb{C}^{N_{\textnormal{T}} \times N_{\textnormal{R}} \times N_{\textnormal{C}}}
    \;\Longrightarrow\;
    \mathbf{X}_{i} \in \mathbb{R}^{N_{\textnormal{T}}N_{\textnormal{R}} \times N_{\textnormal{C}} \times 2},
\end{align}
which preserves both spatial ($N_{\textnormal{T}}N_{\textnormal{R}}$) and subcarrier-domain ($N_{\textnormal{C}}$) structure.  
The four channel tensors are then stacked as
\begin{align}
    \mathbf{X}
    =
    \bigl[ \mathbf{X}_{2},\mathbf{X}_{5},\mathbf{X}_{8},\mathbf{X}_{11} \bigr]
    \in
    \mathbb{R}^{T_u \times N_{\textnormal{T}}N_{\textnormal{R}} \times N_{\textnormal{C}} \times 2},
\end{align}
where $T_u = 4$ denotes the number of UL DM-RS symbols per slot.
Each $\mathbf{X}_{i}$ is partitioned into non-overlapping patches of size $(h_p, w_p)$, yielding
\begin{align}
    P
    =
    \frac{N_{\textnormal{T}} N_{\textnormal{R}}}{h_p}
    \cdot
    \frac{N_{\textnormal{C}}}{w_p},
\end{align}
where $P$ is the total number of patches.  
Let $\mathbf{X}_{i,p} \in \mathbb{R}^{h_p \times w_p \times 2}$ denote the $p$-th patch at time index $i$.
Each patch is flattened and embedded as
\begin{align}
    \mathbf{e}_{i,p}
    =
    \mathbf{W}_{\mathrm{emb}} \,
    \mathrm{vec}(\mathbf{X}_{i,p})
    +
    \mathbf{b}_{\mathrm{emb}}
    +
    \mathbf{p}_{p},
\end{align}
where $\mathbf{W}_{\mathrm{emb}}$ and $\mathbf{b}_{\mathrm{emb}}$ are learnable projection parameters and 
$\mathbf{p}_{p}$ is a positional encoding.  
The resulting token matrix for time index $i$ is
\begin{align}
    \mathbf{Z}_{i}^{(0)}
    =
    \bigl[
        \mathbf{e}_{i,1}^{\top},\dots,\mathbf{e}_{i,P}^{\top}
    \bigr]^{\top}
    \in \mathbb{R}^{P \times D},
\end{align}
where $D$ denotes the token embedding dimension.

\paragraph{Spatial Transformer Encoder}
Each token matrix $\mathbf{Z}_{i}^{(0)} \in \mathbb{R}^{P \times D}$ is processed by
$L_s$ spatial transformer layers.  
Each layer consists of multi-head self-attention (MHSA) and a feedforward network (FFN).  
The attention operation is
\begin{align}
    \mathrm{Attention}(Q,K,V)
    =
    \mathrm{softmax}\!\Bigl( \tfrac{QK^{\top}}{\sqrt{d}} \Bigr)V,
\end{align}
where $Q,K,V$ are linear projections of the input and $d$ is the per-head dimension.
The layer updates are
\begin{align}
    \widetilde{\mathbf{Z}}_{i}^{(\ell)}
    &=
    \mathbf{Z}_{i}^{(\ell-1)}
    +
    \mathrm{MHSA}\!\bigl(\mathrm{LN}(\mathbf{Z}_{i}^{(\ell-1)})\bigr), \\
    \mathbf{Z}_{i}^{(\ell)}
    &=
    \widetilde{\mathbf{Z}}_{i}^{(\ell)}
    +
    \mathrm{FFN}\!\bigl(\mathrm{LN}(\widetilde{\mathbf{Z}}_{i}^{(\ell)})\bigr),
\end{align}
where $\ell = 1,\dots,L_s$ and $\mathrm{LN}(\cdot)$ is layer normalization.

\paragraph{Temporal Transformer Encoder}
The spatially processed tokens $\mathbf{Z}_{i}^{(L_s)}$ 
are averaged across patches to obtain a $D$-dimensional summary:
\begin{align}
    \mathbf{u}_{i}
    =
    \tfrac{1}{P}
    \sum_{p=1}^{P}
        \mathbf{Z}_{i,p}^{(L_s)}
    \in \mathbb{R}^{D},
\end{align}
where $\mathbf{Z}_{i,p}^{(L_s)}$ is the $p$-th row (patch token) of $\mathbf{Z}_{i}^{(L_s)}$.
Stacking the four summaries yields the temporal sequence
\begin{align}
    \mathbf{U}^{(0)}
    =
    \bigl[
        \mathbf{u}_{2}^{\top},\mathbf{u}_{5}^{\top},
        \mathbf{u}_{8}^{\top},\mathbf{u}_{11}^{\top}
    \bigr]^{\top}
    \in \mathbb{R}^{T_u \times D}.
\end{align}
Temporal MHSA and FFN blocks are applied for $L_t$ layers:
\begin{align}
    \widetilde{\mathbf{U}}^{(\ell)}
    &=
    \mathbf{U}^{(\ell-1)}
    +
    \mathrm{MHSA}_t\!\bigl(\mathrm{LN}(\mathbf{U}^{(\ell-1)})\bigr), \\
    \mathbf{U}^{(\ell)}
    &=
    \widetilde{\mathbf{U}}^{(\ell)}
    +
    \mathrm{FFN}_t\!\bigl(\mathrm{LN}(\widetilde{\mathbf{U}}^{(\ell)})\bigr),
\end{align}
where $\mathrm{MHSA}_t$ and $\mathrm{FFN}_t$ operate along the temporal axis.

A slot-level representation is obtained by temporal averaging:
\begin{align}
    \mathbf{h}
    =
    \tfrac{1}{T_u}
    \sum_{i \in \{2,5,8,11\}}
        \mathbf{U}_{i}^{(L_t)}
    \in \mathbb{R}^{D},
\end{align}
and processed by an MLP:
\begin{align}
    \mathbf{o}
    &= 
    \mathbf{W}_{2}\sigma(\mathbf{W}_{1}\mathbf{h}+\mathbf{b}_{1})+\mathbf{b}_{2}, \\
    \boldsymbol{\pi}
    &= 
    \mathrm{softmax}(\mathbf{o}),
\end{align}
where $\boldsymbol{\pi} \in \mathbb{R}^{|\mathcal{P}|}$ is the predicted distribution over allocation patterns.
The final allocation is
\begin{align}
    \hat{p}^{*}
    =
    \arg\max_{p \in \mathcal{P}} \pi_{p}.
\end{align}


The ViViT complexity is $\mathcal{O}(L(N^{2}D + ND^{2}))$, where $L$ is the total number of transformer layers, 
$N = P T_{u}$ the number of input tokens, and $D$ the embedding dimension.
The CNN baseline consists of $L$ 3D convolutional layers with complexity $\mathcal{O}\!\left(L F C K^{3} H W T\right)$, where $C$ and $F$ are the input and output channels, $K$ is the kernel size, and $HWT$ is the spatial–temporal resolution.
Both models achieve microsecond-level inference on modern GPUs, enabling operation within the shortest NR slot (1/64 ms), with further latency reduction possible via quantization~\cite{ZMORA2021}.
\footnote{With $L{=}8$, $N{=}64$, $D{=}128$, the ViViT model requires $\sim1.082\times10^9$ FLOPs, corresponding to $\sim3.47\,\mu$s on an NVIDIA A100 (312~TFLOPs). The CNN model requires fewer FLOPs.}

\subsection{Dataset Generation}

\begin{figure}[ht]
\captionsetup[subfloat]{farskip=2pt}
\vspace{-7pt}
\centering
        \subfloat[Scene]{\includegraphics[width=.44\linewidth]{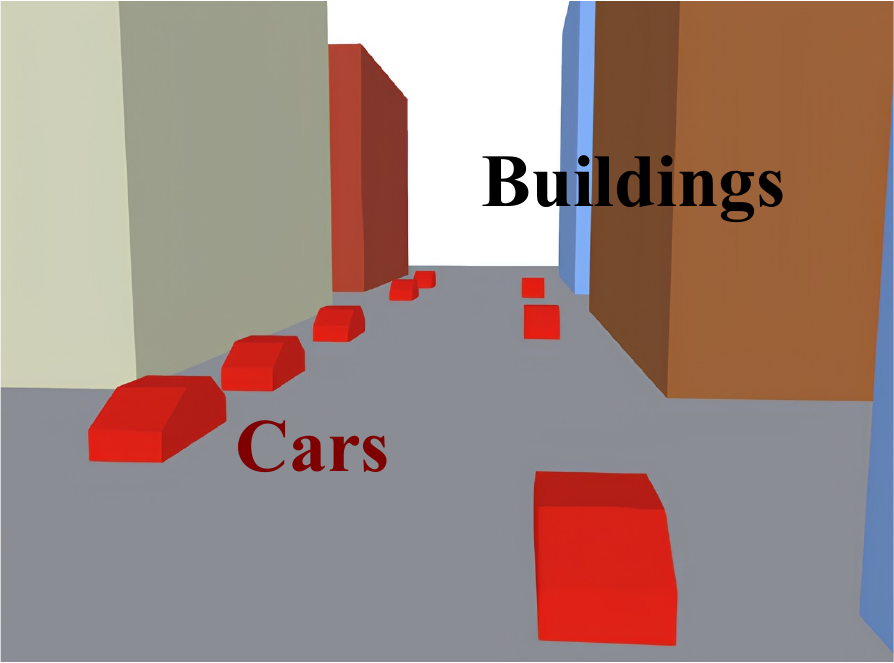}}
        \hspace{0.03\linewidth}
        \subfloat[Ray Tracing]{\includegraphics[width=.44\linewidth]{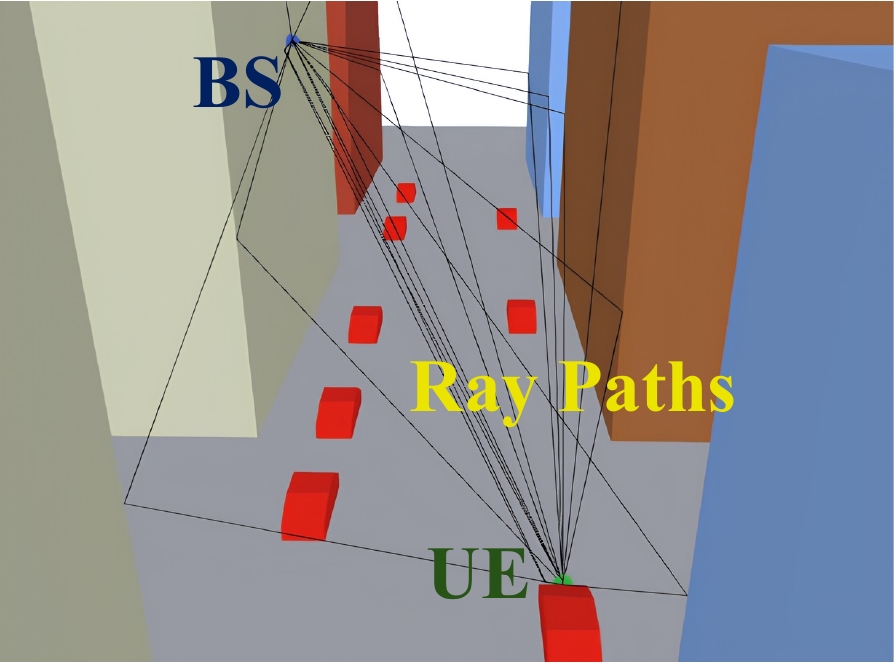}}\\
\vspace{-3pt}
\caption{Process of obtaining channel matrices via ray tracing using NVIDIA Sionna \cite{NVIDIA}. After constructing a dynamic scene that reflects movement as illustrated in (a), perform ray tracing by positioning the BS and UE as shown in (b).}
\vspace{-2pt}
\label{fig:sionna_RT}
\end{figure}

We utilized the ray-tracing module of NVIDIA Sionna to generate channel data \cite{NVIDIA}, as it more accurately reflects real-world propagation characteristics and physical environments than statistical channel models.
We employed the \textit{simple\_street\_canyon\_with\_cars} scene, where vehicles move along a two-lane road between buildings, as shown in Fig. \ref{fig:sionna_RT}.
The BS is mounted on a building, and the UE is placed in a vehicle moving at $v \in \{ 60, 65, \dots, 80 \}$ km/h.
The BS has an 8×8 antenna array, while the UE has a 2×1 antenna array, forming a massive MIMO system.
We consider an FDD scenario operating in NR band n1, as specified in Table 5.2-1 of \cite{TS138101}, using uplink and downlink carrier frequencies of 1.95 GHz and 2.14 GHz, respectively.
The numerology is fixed to N=1, such that one subframe corresponds to a single slot.
The maximum number of rays was set to 20, with a maximum ray depth of 4 and a bandwidth of 100 MHz.


To emulate practical estimation errors, we added Gaussian noise with magnitude $\approx$0.05 relative to the original channels, generating 4,000 additional slots and yielding 6,000 slots in total.
For each slot, the optimal DM-RS allocation was obtained via exhaustive search using Sionna link-level simulations based on ($\ref{eq:optimization_problem}$), and the dataset was split into training/validation/test sets with an 8:1:1 ratio.

\section{Experimental Results}

\vspace{-3pt}
\subsection{Considered Schemes}

\noindent\textbf{CPRS (ViViT)}:
The proposed scheme described in Section~\ref{SEC:CPRS_ViViT}.
It adopts a tubelet embedding with a patch size of $(1,8,8)$ and a Transformer backbone consisting of 8 encoder layers, each with 16 attention heads and an embedding dimension of 128.
These architectural parameters were selected based on ablation studies, where this configuration consistently achieved the best performance among the considered ViViT variants.

\noindent\textbf{CPRS (CNN)}:
A CPRS variant based on a 3D convolutional neural network, where all $\mathbf{X}_{i}$ are concatenated into a four-channel input.
The network consists of three Conv3D layers with 16, 32, and 64 filters, respectively, each followed by batch normalization and 3D max pooling, and a 128-unit dense layer at the output.
This configuration was chosen as it yielded the highest performance among evaluated CNN-based architectures.

\noindent\textbf{Disjoint (ViViT/CNN)}:
In the disjoint baseline, the downlink CSI is first predicted using a ViViT- or CNN-based channel predictor~\cite{YANG2020}.
To ensure a fair comparison, the ViViT predictor adopts the same patch size $(1,8,8)$ and backbone depth (8 layers with 16 attention heads) as CPRS (ViViT), while the CNN predictor uses the same 3D-CNN architecture as CPRS (CNN).
The predicted CSI is then used to optimize the DM-RS pattern by maximizing channel estimation accuracy, following the procedure in~\cite{MASHHADI2021Prun} with adaptations for NR compliance.

\noindent\textbf{DRL}~\cite{KIM2023}:
A CNN-based deep Q-network employing an $\epsilon$-greedy policy ($\epsilon = 0.3$), where the uplink CSI serves as the state, 3GPP-compliant DM-RS patterns constitute the action space, and throughput is used as the reward.

\noindent\textbf{Maximize Data Symbols}:
A heuristic scheme that maximizes the number of transmitted data symbols by minimizing the number of DM-RS symbols per slot ($n_p = 1$).

\noindent\textbf{Random Average}:
A scheme that selects all possible DM-RS allocations with equal probability.
The reported performance corresponds to the average over all possible allocations.

\noindent\textbf{Best}:
The ground-truth DM-RS allocation that achieves the optimal data throughput obtained through simulations.

\begin{table}[t]
\centering
\caption{Classification accuracy comparison of different schemes for selecting the optimal DM-RS allocation index in the next slot, evaluated over 250 test slots. The results are reported per SNR value, excluding SNRs below –5 dB where all DM-RS allocations yield zero data throughput.}
\label{tab:Accuracies}
    \adjustbox{width=0.99\linewidth}
    {
    \begin{tabular}{>{\centering\arraybackslash}m{1.7cm} c c c c c c c c c c}
        \toprule[1pt]
        \multirow{2}{*}[-4pt]{Method} & \multicolumn{9}{c}{SNR (dB)} \\
         \cmidrule[0.5pt](l{1pt}r{1pt}){2-10}
          & -5 & -2.5 & 0 & 2.5 & 5 & 7.5 & 10 & 12.5 & 15\\
        \cmidrule[0.5pt](l{1pt}r{1pt}){1-10}
        \shortstack{CPRS\\(ViVIT)} & \underline{\textbf{99.78}} & \underline{\textbf{99.50}} & \underline{\textbf{100.00}} & \underline{\textbf{99.67}} & \underline{\textbf{99.84}} & \underline{\textbf{99.89}} & \underline{\textbf{100.00}} & \underline{\textbf{99.83}} & \underline{\textbf{99.95}} \\
        \arrayrulecolor{lightgray}
        \cmidrule[0.5pt](l{1pt}r{1pt}){1-10}
        \shortstack{CPRS\\(CNN)} & \underline{94.45} & \underline{97.67} & \underline{\textbf{100.00}} & \underline{99.61} & \underline{\textbf{99.84}} & \underline{99.72} & \underline{99.95} & \underline{99.78} & \underline{\textbf{99.95}} \\
        \cmidrule[0.5pt](l{1pt}r{1pt}){1-10}
        \shortstack{Disjoint\\(ViViT)} & 86.66 & 78.33 & \underline{\textbf{100.00}} & 85.66 & 76.00 & 77.66 & 79.00 & 75.33 & 71.33 \\
        \cmidrule[0.5pt](l{1pt}r{1pt}){1-10}
        \shortstack{Disjoint\\(CNN)} & 86.00 & 77.66 & \underline{\textbf{100.00}} & 83.66 & 72.33 & 69.66 & 70.00 & 68.66 & 67.33 \\
        \cmidrule[0.5pt](l{1pt}r{1pt}){1-10}
        \shortstack{DRL\\(\cite{KIM2023})} & 13.46 & 12.00 & 45.61 & 42.95 & 31.06 & 52.33 & 48.33 & 43.00 & 43.61 \\
        \cmidrule[0.5pt](l{1pt}r{1pt}){1-10}
        \shortstack{Maximize\\Data Symbols} & 75.00 & 72.30 & 75.00 & 50.00 & 50.00 & 31.97 & 25.00 & 25.00 & 25.00 \\
        \cmidrule[0.5pt](l{1pt}r{1pt}){1-10}
        \shortstack{Random\\Average} & 7.69 & 7.69 & 7.69 & 7.69 & 7.69 & 7.69 & 7.69 & 7.69 & 7.69 \\
        \arrayrulecolor{black}
        \bottomrule[1pt]
        \multicolumn{10}{l}{\small*Best: \underline{\textbf{bold}}, second-best: \underline{underline}.}
    \end{tabular}
    }
    \vspace{-9pt}
\end{table}

\vspace{-8pt}
\subsection{Performance Comparison} 
We evaluated each scheme over SNRs from $-10$ to $15$ dB.
As shown in Table \ref{tab:Accuracies}, the proposed algorithm, CPRS (ViViT), achieves the highest classification accuracy across all SNR values, with performance consistently approaching 100\%.
In the disjoint approach, the channel predictor achieves accurate downlink channel estimation with an NMSE of –8.42 dB, and the DM-RS pruning network attains over 98\% test accuracy when provided with perfect downlink CSI.
However, when these two modules are cascaded, error accumulation leads to substantial performance degradation.
Even when similar NMSE levels are achieved, the end-to-end performance varies significantly depending on how well the predicted CSI preserves channel structures that are most relevant for reference signal placement.
The DRL method, requiring a clearer understanding of the complex relationship between uplink CSI and throughput, converged to a lower overall performance compared to CPRS.

\begin{figure}[t]
\vspace{-7pt}
    \centering
    \includegraphics[draft=false, width= 0.99 \linewidth]{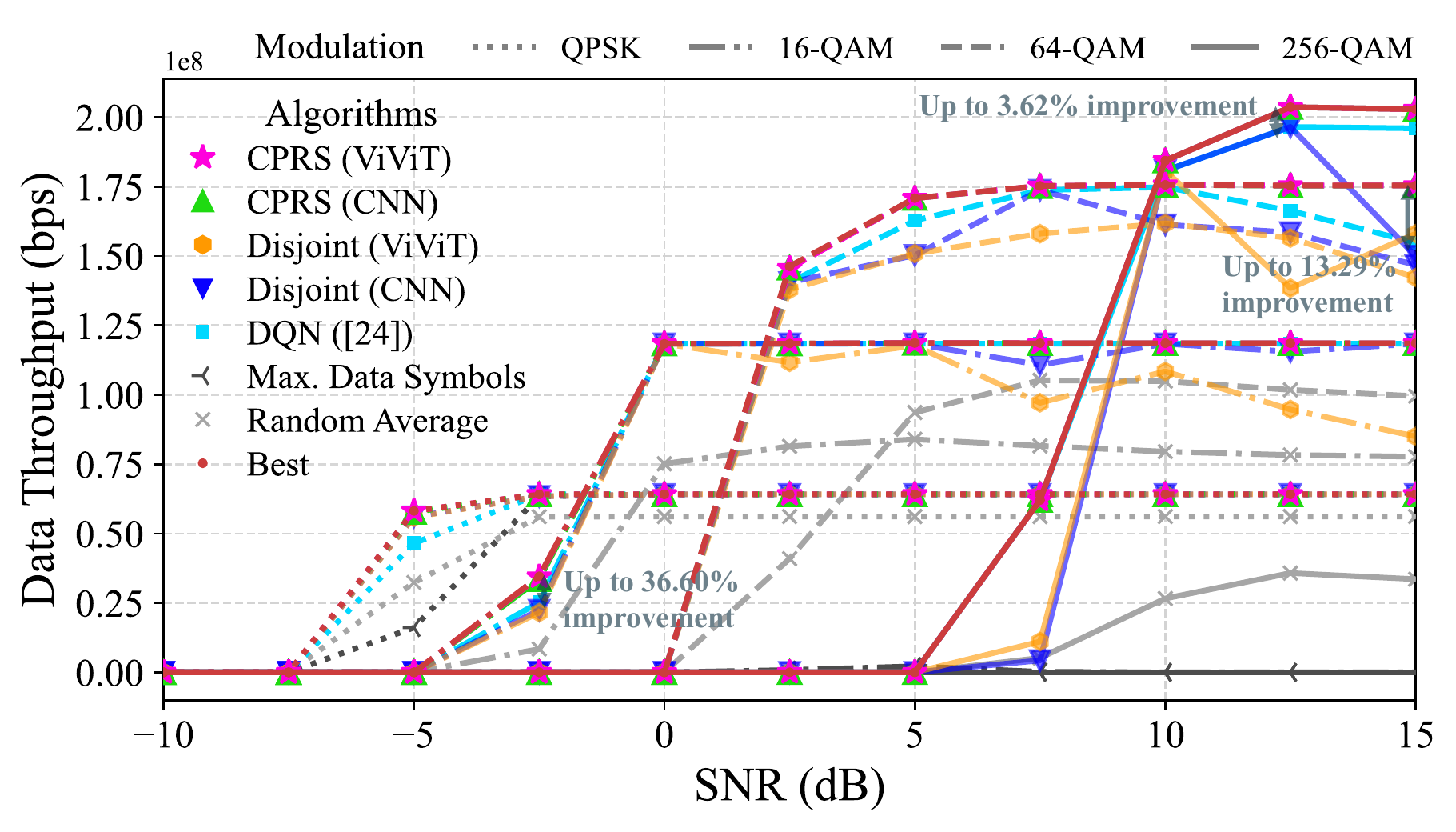}
\vspace{-17pt}
    \caption{Average data throughput achieved by each scheme's DM-RS allocation, evaluated over 250 randomly selected sample slots. The throughput gain of the proposed CPRS algorithm is shown relative to the second-best benchmark.}
    \label{fig:SNR_THR}
    \vspace{-14pt}
\end{figure}

The key performance indicator at the UE is data throughput, as shown for each scheme in Fig.~\ref{fig:SNR_THR}.
CPRS (ViViT) achieves the highest throughput across the entire SNR range and modulation orders, consistently approaching the optimum with a relative error of only 0\%–0.48\% compared to the best achievable performance.
Compared to the second-best benchmark, CPRS (ViViT) provides up to a 36.60\% throughput gain, and up to a 13.29\% gain in the saturation region under 64-QAM.
CPRS (CNN) ranks second overall, with an average throughput that is only 0\%–2.60\% lower than that of CPRS (ViViT), while other baseline methods exhibit degraded performance in fine-grained DM-RS allocation, particularly at higher modulation orders.
More importantly, compared with conventional disjoint approaches, the proposed CPRS framework achieves up to a 459.79\% performance improvement and an average gain of 28.85\%.
These results demonstrate the necessity of the CPRS framework for jointly optimizing channel prediction and DM-RS allocation, rather than treating them as separate modules.



\vspace{-5pt}
\section{Conclusion}

In this paper, we presented channel prediction-based reference signal allocation (CPRS) to reduce CSI feedback overhead in massive MIMO FDD systems.
The proposed ViViT/CNN-based CPRS learns spatio–temporal patterns from uplink CSI to determine the DM-RS allocation and achieves up to a 36.60\% throughput gain over benchmark schemes in Sionna ray-tracing simulations.
This study provides a foundation for extending prediction-driven reference signal optimization beyond DM-RS, including applications to B5G systems and joint design with beamforming under diverse mobility and channel conditions.


\footnotesize
\bibliographystyle{IEEEtran}
\bibliography{ref}

\end{document}